\documentclass[article, eqsecnum, aps]{revtex4}
\begin{document}
\title{ Schwinger Representation for the Symmetric Group: Two explicit constructions for the Carrier Space }

\author{S. Chaturvedi
\\
Institute for Mathematical Sciences, Imperial College London,\\ London SW7
2BW, UK,\\
and\\ School of Physics, University of Hyderabad, \\Hyderabad 500 046,
India \thanks {scsp@uohyd.ernet.in}\\
G. Marmo
\\Dipartimento di Scienze Fisiche, 
\\Universita di Napoli Federico II and
INFN, Via Cintia, 80126 Napoli, Italy \thanks{giuseppe.marmo@na.infn.it}\\
N. Mukunda
\\Centre for High Energy Physics, Indian Institute of Science,\\
Bangalore~560~012, India \thanks{nmukunda@cts.iisc.ernet.in}\\
R. Simon\\ The Institute of
 Mathematical Sciences,\\ C. I. T. Campus, Chennai 600 113, India 
 \thanks{simon@imsc.res.in}}
\begin{abstract}
We give two explicit construction for the carrier space for the Schwinger representation of the group $S_n$. While the first relies on a class of functions consisting of monomials in antisymmetric variables, the second is
 based on the Fock space associated with the Greenberg algebra.
\end{abstract}
\maketitle

\newpage
\section{Introduction}
Given a group $G$ it is desirable to construct a representation of $G$ such that it contains every irreducible representation of $G$ exactly once. Building such economical and complete compendia of irreducible representations for various groups has been and continues to be of great interest in mathematics where they are usually referred to as Gelfand models \cite{bernstein} or simply as models. Way back in the fifties, a model in this sense, for the group $SU(2)$, was introduced by Schwinger \cite{schwinger} making use of the algebra of the annihilation and creation operators of two independent quantum mechanical harmonic oscillators. Since its inception, this Schwinger Oscillator constructiion has found many useful physical applications in a wide variety of contexts which include the physics of strongly correlated systems \cite{arovasaur}, quantum optics of two mode radiation fields \cite{ardutta}, analysis of partially coherent classical Gaussian Schell model beams \cite{sundar}, extension to all three-dimensional Lie algebras and analysis of both classical and q-deformed versions \cite{manko}, applications in the context of quantum computing \cite{ruben}, and a new approach to the spin-statistics theorem \cite{berry}. The concept of a Schwinger Representation, as a model for containing all unitary irreducible representations of compact Lie groups and finite groups was formalised in a recent work \cite{concept}  
and Schwinger representations for $SU(2), SO(3)$ and $SU(n)$ for all $n$  were constructed via specific carrier spaces and group actions. In this work the role of the Schwinger Representation in setting up the Wigner-Weyl isomorphism for quantum mechanics on a compact simple Lie group \cite{mio}  was also highlighted.

The purpose of this work is to develop a Schwinger Representation for the symmetric group $S_n$. Models for $S_n$ \cite{inglis}, \cite{aguado} and $GL(n,q)$ , the general linear group over finite fields, have been explicitly constructed and investigated in detail in mathematical literature \cite{klya},\cite{inglis1},\cite{soto}. These works, while accomplishing their main objective with mathematical elegance and sophistication, by and large, tend not to highlight  aspects pertaining to the the underlying carrier space to the extent one would desire so as to see possible applications of these results to theoretical physics. A much more `direct' construction of a `natural representation' for $S_n$  recently given by Kodiyalam and Verma in \cite{vijay}, comes more in line with a theoretical physicist's approach, in that it spells out the carrier space at the very outset. The key feature which lends this construction its `naturalness' is the observation that the the number of elements of $S_n$ which are involutions i.e. which square to the identity element    exacly equals the sum of the dimensions of the irreducible representations of $S_n$. This set therefore provides the most natural choice for a set on which to build a complete multiplicity free representation of $S_n$. The present work is entirely based on the Kodiyalam-Verma construction and our principal contribution, much in the spirit of the Schwinger's original construction for $SU(2)$, consists in giving their carrier space a more tangible appearance in the form of (a) 
a Hilbert space of functions consisting of monomials in antisymmetric variables (b) a subspace of a Fock space built on the algebra of creation and annihilation opertors proposed by Greenberg\cite{greenberg} .   

A brief outline of this paper is as follows. In Section II we recapitulate basic elements of Kodiyalam-Verma construction and provide necessary motivation and background for our results that follow. In section III and IV we give an explicit construction for the space carrying the Schwinger representation of $S_n$ based on monomials in antisymmetric variables  and that based on the Fock space associated with the Greenberg algebra. Section IV contains our concluding remarks. 

\section{ Elements of Kodiyalam-Verma Construction} 
Consider the group $S_n$, the set of permutations of $n$ objects $\{1,2,\cdots,n\}$. Let $X_m$ with $m$ taking values $1,2\cdots, {\rm Int}[n/2]$  denote the conjugacy class of length $m$
i.e. consisting of products of $m$ distinct transpositions. Denoting a transposition by $(i_1~i_2)$, the members of $X_m$ can be arranged in a standard form as follows:             
\begin{equation}
\{(i_1~i_2)\cdot (i_3~i_4)\cdots (i_{2m-1}~i_{2m})\}
\end{equation}
where $i_1,i_2,\cdots i_{2m}$ take values $1,2,\cdots ,n$ and satisfy
\begin{eqnarray}
 i_1<i_2,~i_3<i_4,&&\cdots~,i_{2m-1}<i_{2m},\nonumber\\
 i_1<i_3<i_5<&&\cdots i_{2m-1},\\
  i_2\neq i_4\neq &&\cdots i_{2m}.\nonumber.
\end{eqnarray}
\noindent
To the $X_m$'s above we append a set $X_0$ taken to be the identity element of $S_n$ and define 
$$X={\bigcup}_{m=0}^{{\rm Int}[n/2]}~~X_m$$
The number of elements in $X$ equals the sum of the dimensions of the irreducible representations of $S_n$. 
Thus for, instance, for $S_4$ we have:
\begin{eqnarray}
X_1&=&\{(1~2),(1~3),(1~4),(2~3),(2~4),(3~4)\}\nonumber \\
X_2&=&\{(1~2)(3~4),(1~3)(2~4),(1~4)(2~3)\}
\end{eqnarray}
which together with $X_0$ correctly add up to $10$. 

Consider now the usual action of elements $\pi$ of $S_n$ on the elements of  $X_m$ : 
\begin{equation}
\pi : (i_1~i_2)\cdot (i_3~i_4)\cdots (i_{2m-1}~i_{2m})\rightarrow
(\pi(i_1)~\pi(i_2))\cdot (\pi(i_3)~\pi(i_4))\cdots (\pi(i_{2m-1})~\pi(i_{2m})),
\end{equation}
rearrange the result into the standard form, and ascribe it a sign 
given by 
\begin{equation}
(-1)^{\#
\{k:(\pi(i_{2k-1})>\pi(i_{2k})\}}
\label{a}
\end{equation}
where $\#$ stands for `cardinality of'. Clearly this action which maps the elements of $X_m$, for each $m$, into themselves upto a sign  leads to a representation of $S_n$. Kodiyalam and Verma prove that 

\begin{itemize}
\item Each  $X_m$ affords a representation of $S_n$ which is
multiplicity free.
\item There is no overlap in the irred rep content of the representations
  afforded by different $X_m$'s .
\item The set $$X={\bigcup}_{m=0}^{{\rm Int}[n/2]}~~X_m$$ gives a complete and multiplicity free rep of $S_n$. 
\end{itemize}

The construction of the carrier space for this complete multiplicity free representation in the work of Kodiyalam and Verma is then achieved by a mathematical decree: Regard the elements of $X$ as a set of orthonormal vectors in a Hilbert space. Our aim in the next two sections is to give a more concrete structure to this somewhat 'contrived' Hilbert space. 
A glance at the sign prescription in $(\ref{a})$ suggests viewing the elements of $X_m$ as products of coordinate differences
\begin{equation}
\{[x(i_1)-x(i_2)]\cdot[x(i_3)-x(i_4)]\cdots [x(i_{2m-1})-x(i_{2m})]\}.
\end{equation}
Though this observation by itself does not help us building the desired carrier space -- the elements of this set are not even algebraically independent, it helps us identfy what the ingredients of the final construction we should be looking for. What we need is products of pairs which commute with each others and change sign when the objects within the 
pair are interchanged and at the same time ensure that such products of pairs are algebraically independent. These observations inspire the following two constructions:

\section{Carrier space for the Schwinger Representation of $S_n$ based on monomials in antisymmetric variables}
This construction makes use of the elementary fact that
\begin{equation}
\int_{-\infty}^{\infty} d\xi e^{-\xi^2} \{1~~{\rm or}~~\xi~~{\rm or}~~\xi^2\}=\sqrt{\pi}
\{1~~{\rm or}~~0~~{\rm or}~~\frac{1}{2}\}.
\label{8}
\end{equation} 
We introduce $n(n-1)/2$ real independent variables $\xi_{jk}=- \xi_{kj},~~j,k=1,\cdots,n$. For the class of functions $f(\mbox{\boldmath $\xi$})$ of these variables specified below, we define the inner product as:
\begin{equation}
(f_1(\mbox{\boldmath $\xi$}),f_2(\mbox{\boldmath $\xi$}))=\frac{1}{\pi^{N(N-1)/4}}\int_{-\infty}^{\infty}\cdots \int_{-\infty}^{\infty}
\left(\prod_{j<k} d\xi_{jk}\right)
 e^{-\frac{1}{2}\sum_{j,k}\xi_{jk}^2}~f_1^*(\mbox{\boldmath $\xi$})f_2(\mbox{\boldmath $\xi$}).
\end{equation}
Consider monomials for $r=0,1,2,\cdots,[n/2]$:
\begin{equation}
M_r(\mbox{\boldmath $\xi$})=\xi_{j_1k_1}\xi_{j_2k_2}\cdots\xi_{j_rk_r},
\end{equation}
with 
\begin{eqnarray}
&&j_1,k_1,j_2,k_2\cdots,j_r,k_r~~{\rm all~distinct},\nonumber\\
&&j_1<k_1,j_2<k_2\cdots,j_r<k_r,\\
&&j_1<j_2<j_3<\cdots<j_r.\nonumber
\end{eqnarray}
We consider only $f(\mbox{\boldmath $\xi$})$ which are complex linear combinations of such monomials. Clearly 
\begin{equation}
(M_r(\mbox{\boldmath $\xi$}),M_r(\mbox{\boldmath $\xi$}))=2^{-r}.
\end{equation}
If we take two unequal monomials $M_r(\mbox{\boldmath $\xi$})$,$M_{r^\prime}(\mbox{\boldmath $\xi^\prime$})$, then some factor $\xi_{jk}$ present in $M_r(\mbox{\boldmath $\xi$})$ is absent in $M_{r^\prime}(\mbox{\boldmath $\xi^\prime$})$
or conversely. Thus by $(\ref{8})$, distinct monomials are orthogonal and hence $2^{r/2}
M_r(\mbox{\boldmath $\xi$})$ form an orthonormal set. The symmetric group $S_n$ acting naturally on each  $M_r(\mbox{\boldmath $\xi$})$ produces some $M_r^\prime(\mbox{\boldmath $\xi$})$, with the same $r$ together with signs as required in Kodiyalam-Verma construction. Thus the set $\{2^{r/2}
M_r(\mbox{\boldmath $\xi$}), r=1,\cdots,[n/2]\}$ provides an orthonormal basis for the Hilbert space carrying the Schwinger representation of $S_n$.  
\section{A Carrier Space for the Schwinger Representation for $S_n$ based on Greenberg algebra} 
Our second explicit construction of the carrier space for the the Schwinger representation of $S_n$ is based on the Fock space associated with Greenberg Algebra.
\cite{greenberg}. 

Greenberg, in the conext of Fock spaces for infinite statistics, a second quantised description for the uncorrected Boltzmann statistics, proposed the following `commutation´ relations:   
\begin{equation}
a(i_j)a^\dagger(i_k)=\delta_{i_j,i_k}.
\end{equation}
A peculiar feature of this algebra, which sets it apart from Bose or Fermi algebra is that a string of products of creation ( annihilation) bears no relation to its permuted version-they are not just independent but          'orthogonal' to each other. Thus, for instance, the two particle states   
$a^\dagger(1)a^\dagger(2)|0>$ and $a^\dagger(2)a^\dagger(1)|0>$, which in the Bose and Fermi case describe the same state, in this statistics correspond to two distinct orthogonal states. This special feature of the Greenberg algebra provides us the necessary framework for setting up the carrier space for the Schwinger Representation for $S_n$.

The discussions in the previous sections suggests the following procedure for the desired carrier space:

\begin{itemize}
\item Identify  $X_0$ with the vacuum state $|0>$.
\item Replace the coordinate differences $[x(i_1)-x(i_2)]$ (or $\xi_{i_1i_2})$) by commutators $[a^\dagger(i_1),a^\dagger(i_2)]$ and
\item Replace the products of the coordinate differences ( products of $\xi$)'s by symmetrised products of appropriate commutators  acting on $|0>$ .  
\end{itemize}

The substitute for the sets $X_m$ above now become: 

\begin{equation}
\{\frac{1}{\sqrt{2^m}}~{\bf S}\left[[a^\dagger(i_1),a^\dagger(i_2)]\cdot [a^\dagger(i_2),a^\dagger(i_3)]\cdots [a^\dagger(i_{2m-1}),a^\dagger(i_{2m})]\right]|0>\},
\end{equation}
where $i_1,i_2,\cdots i_{2m}$ take values $1,2,\cdots ,n$ and satisfy
\begin{eqnarray}
 i_1<i_2,~i_3<i_4,&&\cdots~,i_{2m-1}<i_{2m},\nonumber\\
 i_1<i_3<i_5<&&\cdots i_{2m-1},\nonumber\\
  i_2\neq i_4\neq &&\cdots i_{2m}.
\end{eqnarray}
Here ${\bf S}$ indicates symmetrisation of the $m$ factors that appear in the product. 

Thus for instance, for $S_4$, the orthonormal basis carrying the Schwinger representation can be given explicitly as:
\begin{eqnarray}
X_0 &:& |0>. \\
X_1 &:& \frac{1}{\sqrt{2}}~[a^\dagger(1),a^\dagger(2)]|0>,  \frac{1}{\sqrt{2}}[a^\dagger(1),a^\dagger(3)]|0>, \frac{1}{\sqrt{2}}[a^\dagger(1),a^\dagger(4)]|0>\nonumber\\ &&\frac{1}{\sqrt{2}}[a^\dagger(2),a^\dagger(3)]|0>,
\frac{1}{\sqrt{2}}[a^\dagger(2),a^\dagger(4)]|0>,
\frac{1}{\sqrt{2}}[a^\dagger(3),a^\dagger(4)]|0>.\\
X_2 &:&\frac{1}{\sqrt{8}}~\left[[a^\dagger(1),a^\dagger(2)][a^\dagger(3),a^\dagger(4)]+a^\dagger(3),a^\dagger(4)]\right][a^\dagger(1),a^\dagger(2)]|0> ,\nonumber, \\
&&\frac{1}{\sqrt{8}}~\left[[a^\dagger(1),a^\dagger(3)][a^\dagger(2),a^\dagger(4)]+[a^\dagger(2),a^\dagger(4)][a^\dagger(1),a^\dagger(3)]\right]|0> ,\nonumber \\
&&\frac{1}{\sqrt{8}}~\left[[a^\dagger(1),a^\dagger(4)][a^\dagger(2),a^\dagger(3)]+[a^\dagger(2),a^\dagger(3)][a^\dagger(1),a^\dagger(4)]\right]|0>. 
\end{eqnarray}
  
For any $n$, one can also write down the irreducible  representation content of the representation afforded 
by each $X_m$ following a simple recipe. The results upto $m=4$ are given below:
\medskip

\noindent
{\bf Level 0}
$(n)$
\\

\noindent
{\bf Level 1}
\hspace{0.6cm}$(n-1,1)$\hspace{0.1cm}  $(n-2,1^2)$ 
\\

\noindent
{\bf Level 2}
\hspace{2.2cm}$(n-2,2)$  $(n-3,2,1)$ \hspace{0.1cm}$(n-4,2^2)$
\\
\noindent

\hspace{4.8cm}$(n-3,1^3)$\hspace{0.3cm}$(n-4,1^4)$ 
\\

\noindent
{\bf Level 3} 
\hspace{3.8cm}$(n-3,3)$\hspace{0.3cm}  $(n-4,3,1)$ \hspace{0.3cm}$(n-5,3,2)$
\hspace{0.3cm}$(n-6,3^2)$
\\
\noindent

\hspace{6.7cm}$(n-4,2,1^2)$\hspace{0.1cm}  $(n-5,2^2,1)$ \hspace{0.1cm}
$(n-6,2^2,1^2)$
\\
\noindent

\hspace{8.7cm} $(n-5,2,1^3)$\hspace{0.2cm} $(n-6,1^6)$
\\
\noindent

\hspace{8.7cm} $(n-5,1^5)$
\\

\noindent
{\bf Level 4} 
\hspace{5.6cm}$(n-4,4)$\hspace{0.6cm}  $(n-5,4,1)$ \hspace{0.4cm}$(n-6,4,2)$
\hspace{0.3cm}$(n-7,4,3)$\hspace{0.6cm}$(n-8,4^2)$
\\
\noindent

\hspace{8.8cm}$(n-5,3,1^2)$\hspace{0.2cm}  $(n-6,3,2,1)$ 
$(n-7,3^2,1)$\hspace{0.4cm}$(n-8,2^4)$
\\
\noindent

\hspace{10.9cm} $(n-6,2^3)$\hspace{0.5cm} $(n-7,3,2,1^2)$
$(n-8,2^2,1^4)$
\\
\noindent

\hspace{10.9cm} $(n-6,3,1^3)$\hspace{0.1cm} $(n-7,2^2,1^3)$
\hspace{0.2cm}$(n-8,3^2,1^2)$
\\
\noindent

\hspace{10.9cm} $(n-6,1^6)$\hspace{0.4cm} $(n-7,2,1^5)$
\hspace{0.4cm}$(n-8,1^8)$
\\
\noindent

\hspace{13.0cm} $(n-7,1^7)$

The recipe for constructing the above table is as follows:
\begin{itemize}
\item The table is arranged so that all partitions of $n$ of the type 
$(n-i,\cdots)$ appear in the same column. 
\item The first row at each level $m$ involves partitions of the type $(n-m,\cdots),
\cdots(n-m-1,\cdots),\cdots,(n-2m,\cdots)$. Thus $m^{th}$ level has $m+1$ columns.  
\item
The only entry in the first column at every level is $(n-m,m)$. The last 
column contains partitions with the structure, $(n-2m, ~{\rm partitions~of}$ 
$2m$~${\rm  with~ parts~ having~ even~ multiplicities})$. The remaining $m-1$ columns in between at $m^{th}$ level can be obtained by those at the previous level as follows:  
The second column at the $m^{th}$ level is 
obtained by decreasing the first part by $1$ and adding $1$ in all possible 
ways to the remaining parts of the partitions of 
the second column at the previous level such that the result is a partition and 
retaining only the ones that have not already appeared above it. Similarly for the third column and so on.  
\end{itemize} 
Note that each level, the partitions appearing are conjugate to those given by the Inglis-Richardson-Saxl construction at the same level --  $m^{th}$ level contains partitions whose conjugate partitions have $n-2m$ odd parts.

\section{conclusion}
We have given two explicit constructions for the carrier spaces for the Schwinger 
representation for the symmetric group. In light of the ubiquity of the symmetric group in various areas of mathematical physics, we hope that the constructions given here will 
find useful applications. 
\vskip2cm

\noindent
{\bf Acknowledgements:}

\noindent
 One of us (SC)  wishes to thank the Leverhulme
Trust for a Visiting Professorship at the Imperial College London.


\begin{thebibliography}{99}
\bibitem{bernstein} I.N. Bernstein, I.M. Gelfand, S.I. Gelfand, Funct. Anal. Appl.
{\bf 9}, 322 (1975); I.M. Gelfand and A.V. Zelevinskii,
Funct. Anal. Appl. {\bf 18}, 183 (1984).

\bibitem{schwinger} J. Schwinger, {\it On Angular Momentum}, USAEC Report
NYO-3071 (1952); reprinted in L.C. Biedenharn and H. van Dam,
(editors), {\it Quantum Theory of Angular Momentum}, Academic
Press, New York (1965), p.229 and also in Kimball A. Milton
(editor), {\it A Quantum Legacy - Seminal Papers of Julian
Schwinger}, World Scientific Publishing Company, Singapore (2000),
p.173.

\bibitem{arovasaur} D.P. Arovas and A. Auerbach, Phys. Rev. B{\bf 38}, 316
(1988); A. Auerbach and D.P. Arovas, Phys. Rev. Lett.
{\bf 61},  617 (1988); A. Auerbach, {\it Interacting
Electrons and Quantum Magnetism}, Springer, New York (1994).

\bibitem{ardutta} Arvind, B. Dutta, N. Mukunda and R. Simon, Phys. Rev.
A{\bf 52}, 1609 (1993).

\bibitem{sundar} K. Sundar, N. Mukunda and R. Simon, J. Opt. Soc. Am.
A{\bf 12}, 560 (1995); R. Simon, K. Sundar, N. Mukunda, J. Opt.
Soc. Am. A{\bf 10}, 2008 (1993).

\bibitem{manko} V.I. Man'ko, G. Marmo, P. Vitale and F. Zaccaria,
 Int. J. Mod. Phys. A{\bf 9}, 5541 (1994).

\bibitem{ruben} P. Aniello and R. Coen Cagli, J.Opt.B Quant.Semiclass.Opt. {\bf 7}, S711(2005).
\bibitem{berry} M.V. Berry and J.M. Robbins, Proc. Roy. Soc. London,
A{\bf 453}, 1771 (1997).
\bibitem{concept} S. Chaturvedi, G. Marmo, N. Mukunda, R. Simon and A. Zampini, Rev. Math. Phys. {\bf 18}, 887 (2006). 

\bibitem{mio} N. Mukunda, G. Marmo, A. Zampini, S. Chaturvedi and R. Simon,
 J.Math. Phys. {\bf 46}, 012106 (2005).


\bibitem{inglis} N.F.J. Inglis, R.W. Richardson and J. Saxl,
Arch. Math. {\bf 54}, 258 (1990).
\bibitem{aguado} J.L. Aguado and J.O. Araujo, Comm. Alg., {\bf 29}, 1841 (2001).
\bibitem{inglis1} N.F.J. Inglis and J. Saxl,
Arch. Math. {\bf 57}, 424 (1991).
\bibitem{klya} A.A. Klyachko, Math. of the USSR-Sbornik, {\bf 48}, 365 (1984).
\bibitem{soto} J. Soto-Andrade {\it Geometrical Gelfand models, tensor quotients and Weil representations}, in Proceedings of Symposia in Pure Mathematics, Vol. 47, part 2, 306 (1987). 

\bibitem{vijay} V. Kodiyalam and D.-N. Verma, {\it A natural representation model for symmetric group} IMSc preprint, arXiv:math.RT/0402216v1.

\bibitem{greenberg}
O.W. Greenberg, Phys. Rev. Lett. {\bf 64}, 705 (1990).

\end{thebibliography}
\end{document}